\documentclass[pre,reprint]{revtex4-1}

\newcommand{\rmd}{\ensuremath{\mathrm{d}}}
\newcommand{\arsinh}{\ensuremath{\mathrm{arsinh}}}

\newcommand{\abs}[1]{\ensuremath{\left\vert#1\right\vert}}
\usepackage[lofdepth,lotdepth,caption=false]{subfig}
\usepackage{amsmath}
\usepackage{amssymb}
\usepackage{mathtools}
\usepackage{graphicx}
\usepackage{dcolumn}
\usepackage{color}

\begin{document}

\title{Kinetic modeling of plasma destruction by a dielectric solid}
\title{Kinetic modeling of the electric double layer at a dielectric plasma-solid interface}
\author{K. Rasek, F. X. Bronold and H. Fehske}
\date{\today}

\address{Institut f{\"u}r Physik,
	Universit{\"a}t Greifswald, 17489 Greifswald, Germany }

\begin{abstract}
For a collisionless plasma in contact with a dielectric surface,
where with unit probability electrons and ions are, respectively,
absorbed and neutralized, thereby injecting electrons 
and holes into the conduction and valence band, we study the kinetics
of plasma loss by nonradiative electron-hole recombination inside 
the dielectric. We obtain a self-consistently embedded electric double 
layer, merging with the quasi-neutral, field-free regions inside
the plasma and the solid. After a description of the numerical scheme 
for solving the two sets of Boltzmann equations, one for the electrons 
and ions of the plasma and one for the electrons and holes of the 
solid, to which this transport problem gives rise
to, we present numerical results for a p-doped dielectric. Besides
potential, density, and flux profiles, plasma-induced changes in the 
electron and hole distribution functions are discussed, from 
which a microscopic view on plasma loss inside the dielectric emerges.
\end{abstract}

\maketitle

\section{Introduction}
At the interface between a low-temperature plasma and a macroscopic solid an 
electric double layer forms consisting of a plasma-bound electron-depleted  
and a solid-bound electron-rich region. In the simplest scenario, 
the charge separation arises because electrons, outrunning ions on the plasma 
side, are more efficiently deposited into the surface than they are extracted 
from it by the neutralization of ions which effectively leads to the injection 
of missing electrons, that is, in the language of solid state physics to the 
injection of holes. At the end a potential profile builds up equalizing the electron 
and ion fluxes issued by the plasma source with the electron-hole recombination
flux inside the solid. The double layer is hence caused by the plasma but 
controlled by the solid. 

Little is quantitatively known about the scenario although the positive part 
of the double layer--the plasma sheath--has been studied in great detail 
ever since the work by Langmuir and Mott-Smith~\cite{LM24}. Most of 
the studies focus on the merging of the sheath with the quasi-neutral 
bulk plasma~\cite{SB90,Riemann91,Franklin03,Brinkmann09,Robertson13}. The 
effect of the solid is studied only in as far as its emissive properties, 
electron/ion reflection and secondary electron emission, affect the stability 
of the sheath~\cite{HZ66,TLC04,SKR09,SHK13,LW15,CU16}. The reasoning behind 
it is the assumption that processes inside the solid occur on spatio-temporal 
scales too small or too fast to affect the physics of the plasma~\cite{Franklin76}. 
For the plasma species the solid is thus only a sink or source characterized by 
probabilities for absorption, reflection and emission which, in principle, can be 
measured~\cite{MCA15,DAK15,DBS16} or calculated~\cite{BF15,PBF18,DDM19}. There are 
however also theoretical approaches~\cite{HBF12,BF17,BFA19,APA20} treating the 
solid and the plasma as two sides of an interface to be analyzed together.

Mapping the charge dynamics of the solid in contact with the plasma to a set 
of parameters is no longer justified in situations where the scales of 
the plasma and the solid become comparable or where the solid is an integral part
of the plasma device of interest as it is, for instance, the case in attempts
to combine gaseous with solid state electronics~\cite{OE05,DOL10,TWH11,KSO12,TP17}. 
In particular, if the miniaturization of the devices continues~\cite{EPC13}, the 
transit times through the plasma and the transport and relaxation times inside the 
solid may become comparable, requiring then to resolve the charge dynamics inside 
the solid and the plasma at the same kinetic level. 
 
Recently, we set up a theoretical framework showing how such a calculation can be 
organized for a plasma-facing dielectric solid~\cite{BF17}. It is based on two
sets of spatially separated Boltzmann equations, one for the electrons and ions 
inside the plasma and one for the conduction band electrons and valence band holes 
inside the dielectric. The two sets are coupled by the electric field, entering the 
force terms of the Boltzmann equations and being the solution of the Poisson equation,
and matching conditions at the interface describing electron transmission and
reflection in either way as well as hole injection due to the neutralization of ions. 
To demonstrate the feasibility of the approach, we applied it to a collisionless, 
perfectly absorbing interface with an ad-hoc recombination condition to 
prevent--in a collisionless situation--the unlimited growth of the charge 
carriers inside the solid. Although conceptually incomplete at this point, 
it seemed useful because a numerical solution of the Boltzmann equation could 
be avoided. 

The purpose of the present work is to remedy this shortcoming by applying the 
theoretical framework to an interface which is left collisionless only on the 
plasma side, where it can be justified, because electrons are strongly depleted,
scattering hence only weakly, while ions collide predominantly with neutrals, which
is important only in particular situations~\cite{SG91,Riemann03,Sternovsky05}. 
But by including collisions on the solid side, we can now couple the creation of 
charge carriers by the plasma source to the physical process destroying them 
inside the solid. It is the balancing of the two at quasi-stationarity which 
determines quantitatively the charge and potential profiles on both sides of 
the interface.

The paper is structured as follows. In section~\ref{sec:theory}, divided 
into two subsections, we present in~\ref{subsec:model} a simplified kinetic 
model for the double layer at a dielectric plasma-solid interface and 
in~\ref{subsec:numerics} the numerical strategy for its solution. Energy and 
momentum relaxation due to scattering on optical phonons~\cite{Ridley99} and 
nonradiative electron-hole recombinations due to traps in the energy gap along 
the lines of a kinetic version~\cite{RJH16} of the Shockley-Read-Hall model~\cite{Hall51,SR52} 
are taken into account by the kinetic equations while charge injection is
treated phenomenologically by source functions entering the boundary conditions. 
The numerical approach utilizes an idea of Grinberg and Luryi~\cite{GL92} for 
solving iteratively Boltzmann equations with distribution functions known at the 
two end points of the integration domain, successfully applied to solid-solid 
interfaces~\cite{DG94,DP98,KH02}. Its utility in the present context is based on 
the observation that at the interface the distribution functions can be assumed 
to be known from the previous iteration loop and successively updated until convergence
is reached. Combined with the boundary conditions fixing the distribution functions 
deep inside the solid and the plasma, a transport problem arises to which the 
Grinberg-Luryi approach can be applied in each half-space. Care is however 
required for treating singular points arising either from 
turning points or the vanishing of the electric field due to the embedding between 
field-free bulk regions. Numerical results are given in section~\ref{sec:results} for 
a p-doped dielectric. Potential, charge density, and flux profiles are shown together 
with the distribution functions for the dielectric's surplus carriers 
originating from the plasma. Section~\ref{sec:conclusion} concludes the presentation
and mathematical details are provided in three appendices.

\section{Theory}\label{sec:theory}

\begin{figure}[t]
        \centering
        \includegraphics[width=\linewidth]{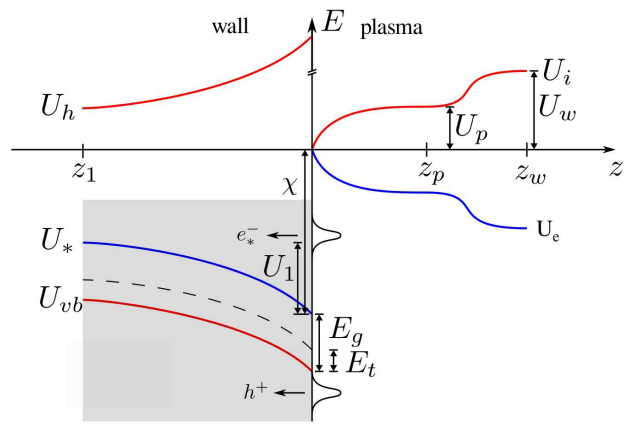}
        \caption{Illustration of the potential energy profiles for an electric 
        double layer at a floating dielectric plasma-wall interface (not on scale). 
        Shown are the edges of the conduction 
        ($U_*$) and valence ($U_{vb}$) bands, the edge for the motion of valence
        band holes ($U_h$), the position of the trap levels ($E_t$), and the 
        potential energy for electrons $(U_e$) and ions ($U_i$) on the plasma side.
        The origin of the energy axis is the potential just out-side the solid,
        $\chi$ is the electron affinity, and $E_g$ the energy gap. The positions
        $z_1$ and $z_p$ are the end points of the double layer and $z_w$ is the 
        location of the plasma source. Also shown are the potential energies at 
        these positions, playing an important role in the modeling. As explained
        in the main text, we cannot resolve the energy space up to the energies 
        where carriers are actually injected from the plasma into the solid. In 
        the simplified model for the solid side, shown in dark, we inject carriers 
        at artificially low energies by source functions included in the matching 
        conditions at $z=0$. 
}
        \label{fig:overview}
\end{figure}

The notation used for the description of the electric double layer at a floating
dielectric plasma-solid interface is summarized in Fig.~\ref{fig:overview}. Also 
shown is the simplification required due to numerical constraints, forcing us to 
restrict the modeling on the solid side to the region close to the band edges. The 
injection of charge carriers into the solid has thus to be taken into account by 
phenomenological source functions.

\subsection{Formulation of the transport problem}\label{subsec:model}

Within the coordinate system of Fig.~\ref{fig:overview}, the plasma-solid interface is 
located at $z=0$ with the solid and plasma filling up the halfspaces $z<0$ and 
$z>0$. The interface is abrupt with material parameters constant and isotropic within 
each halfspace. The spatial dependencies arise from the electric potential energy
$U_c(z)$ for which we set $U_c(0)=0$. The merging of the double layer with the
quasi-neutral, field-free regions occurs at $z_1$ on the solid side and at $z_p$ 
on the plasma side. Since the solid and the plasma accumulate net negative and 
positive charge, respectively, $U_c(z)$ is monotonously increasing with $z$. 

Instead of $z$ we can thus use $U_c$ to track the spatial dependency of all 
physical quantities, with the mapping between the two given by the once-integrated 
Poisson equation,  
\begin{equation}
\label{eq:U'}
\frac{\rmd U_c}{\rmd z} = 
\left(\frac{16 \pi}{\varepsilon(z)}\int\limits_{U_0(z)}^{U_c(z)}\rmd U 
  n(U)\right)^{1/2} 
= \mathcal{E}(U_c)~,
\end{equation}
where $\varepsilon(z)=\Theta(z) + \varepsilon\,\Theta(-z)$, with $\Theta(z)$ the 
usual step function, is the dielectric function of the interface. The function 
$U_0(z)=U_p\Theta(z)+U_1\Theta(-z)$ denotes the potential energy, where 
the net charge vanishes, that is, where quasi-neutrality holds. On the solid
side, this is at $z=z_1$ leading to $U_c=U_1$, where $U_1<0$ is the band bending, 
while on the plasma side it occurs at $z=z_p$ and hence at $U_c=U_p>0$. The 
function $\mathcal{E}(z)$ is the (negative) electric field for which 
\begin{equation}
        \label{eq:matching}
        \varepsilon \mathcal{E} (0^-) = \mathcal{E}(0^+)
\end{equation}
holds at the interface and the (negative) total charge to be integrated over reads 
for a p-doped interface
\begin{multline}
\label{eq:ntot}
	n(U_c) = \left[n_e(U_c) -  n_i(U_c)\right]\Theta(U_c)\\
	+ \left[n_*(U_c) - n_h(U_c) + n_A \right] \Theta(-U_c)~,
\end{multline}
where $n_A$ is the concentration of the acceptors.  

Introducing a species index $s\in \{i,e,h,*\}$ for ions, electrons, valence band holes,
and conduction band electrons, the potential energy $U_s$ can be defined
for each species. Its relation to $U_c$ is given by the following expressions, taking
into account the energy offsets of Fig.\ref{fig:overview}: $U_i = U_c, U_e = -U_c, 
U_h = U_c + E_g + \chi$, and $U_*= -U_c-\chi$, where $\chi$ is the electron affinity
and $E_g$ is the band gap of the dielectric. We will use $U_s$ as a variable synonymous 
to $U_s(U_c)$. 

It is advantageous to introduce in the coordinate system of Fig.~\ref{fig:overview} 
separate distribution functions, $F_s^<$ and $F_s^>$, for the left- and right-moving 
particles with the sign of the perpendicular momentum $k_z$ encoded in the superscript.
Since the interface is homogeneous in the lateral directions it is also rotationally 
invariant in the plane perpendicular to the $z$ axis. The distribution functions 
depend thus only on the magnitude of the lateral momentum ${\bf K}$. Instead of 
it, we use the lateral kinetic energy $T=\hbar^2{\bf K}^2/2m_s$ as a variable, where 
$m_s$ is the mass of a particle of species $s$. In atomic units, measuring 
length in Bohr radii, energy in Rydbergs, and mass in electron masses, the Boltzmann
equation can be cast into
\begin{equation}
\label{eq:BE}
\pm v_s(U_c,E,T)\mathcal{E}(U_c)\frac{\partial}{\partial U_c}F^\gtrless_s(U_c,E,T) = I_\mathrm{coll}^\gtrless~,
\end{equation}
where $k_z$ is replaced by the total energy $E$, $I_\mathrm{coll}^\gtrless$ is the collision 
integral, and
\begin{equation}
\label{eq:vs}
v_s(U_c,E,T) = 2\sqrt{m_s^{-1}(E - U_s-T)}~
\end{equation}
is the velocity perpendicular to the interface. Due to the variable 
transformation~\eqref{eq:ntot} from $z$ to $U_c$ the force term in~\eqref{eq:BE} accounts 
automatically for the Poisson equation. 

The collision integral $I^\gtrless_\mathrm{coll}$, describing scattering and 
recombination processes, depends on either side of the interface on the distribution 
functions of both species. It can be separated into in- and out-scattering parts, 
$\Phi_s^\gtrless$ and $\gamma_s^\gtrless F_s^\gtrless$, respectively, turning the 
Boltzmann equation~\eqref{eq:BE} into its final form,
\begin{equation}
\label{eq:BEQ}
\pm v_s(U_c) \mathcal{E}(U_c) \frac{\partial}{\partial U_c} F^\gtrless_s(U_c) = 
\Phi^\gtrless_s(U_c) - \gamma_s^\gtrless(U_c) F^\gtrless_s(U_c)~,
\end{equation}
where we have omitted the dependencies on $E$, $T$, and $F_s^\gtrless$. In 
appendix~\ref{app:coll} we give $\Phi_s^\gtrless$ and $\gamma_s^\gtrless$ for 
scattering on optical phonons~\cite{Ridley99} and recombination via traps 
in the energy gap~\cite{RJH16}, which is a kinetic formulation of the 
Shockley-Read-Hall model~\cite{Hall51,SR52}. 

Once the solutions of~\eqref{eq:BEQ} are known, the densities
\begin{equation}
\label{eq:n}
n_s(U_c) = \frac{m_s}{8\pi^2}\int\mathrm{d} E \mathrm{d} T \frac{F_s^>(U_c,E,T) + F_s^<(U_c,E,T)}{v_s(U_c,E,T)}~,
\end{equation}
can be obtained, from which the electric field $\mathcal{E}(U_c)$ 
follows by iterating~\eqref{eq:U'}, closing thereby the set of equations.

An essential part of the transport problem are the boundary conditions 
at $U_c=U_1$ and $U_c=U_w$ and the matching condition at $U_c=0$. The 
boundary conditions are given by
\begin{align}
        F_s^> (U_1) = F_s^\mathrm{LM}(U_1)\hspace{30pt}\text{for }&s = h,*~,\\
        F_s^< (U_w) = F_s^\mathrm{LM}(U_w)\hspace{30pt}\text{for }&s = i,e~,
\end{align}
with  
\begin{equation}
\label{eq:LM}
F_s^\mathrm{LM}(U_c) = n_s^\mathrm{LM}\left( \frac{4\pi}{k_B T_s m_s}\right)^{3/2} \exp\left( -\frac{E-U_s}{k_BT_s} \right)~
\end{equation}
a half-Maxwellian with temperature $T_s$ and density $n_s^\mathrm{LM}$~.

The general matching conditions for the distribution functions at $U_c=0$ 
are given in Ref.~\cite{BF17}. We specialize them now to an interface,
where electrons can pass the interface only from the plasma side and ions
are neutralized at the interface with unit probability. Carriers approaching the 
interface from the solid side are specularly reflected. 

Anticipating the potential energy profile of a double layer with negative and positive 
net charge inside the solid and the plasmas, respectively, the matching conditions for 
the electron distribution functions read 
\begin{align}
        F_e^>(0,E,T) &= 0 ~\mathrm{for}~ E>0,\label{eq:BCeI}\\
        F_e^>(U_c,E,T) &= F_e^<(U_c,E,T) ~\mathrm{for}~ E=U_e, \\
        F_*^<(0,E,T) &= F_*^>(0,E,T) + S_*^<(0,E,T)~,\label{eq:BC*I}
\end{align}
while for the ion and hole distribution functions they become 
\begin{align}
        F_i^>(0,E,T) &= 0~,\label{eq:BCiI}\\
        F_h^<(0,E,T) &= F_h^>(0,E,T) + S_h^<(0,E,T)~,\label{eq:BChI}
\end{align}
where we introduced source functions encoding electron and hole injection,
\begin{multline}
\label{eq:Fin}
        S_s^< (U_s,E,T)= n_s^\mathrm{in}\left( \frac{4\pi}{k_B T_s m_s}\right)^{\frac32} \\ 
        \times \exp\left(-\frac{(E-U_s-I_s^\mathrm{in})^2-T^2}{\Gamma_\mathrm{in}^2}\right)~,
\end{multline}
with injection densities $n_s^{\rm in}$ chosen such that $j_h^{\rm in}=j_i$ and 
$j_*^{\rm in}=j_e$. The electron and ion fluxes from the plasma, $j_e$ and $j_i$, are 
given by
\begin{equation}
\label{eq:js}
j_s(U_c) = m_s\int\frac{ \mathrm{d} E\mathrm{d} T}{8\pi^2}\left[F_s^>(U_c,E,T) -F_s^<(U_c,E,T)\right]~,
\end{equation}
and $j_s^{\rm in}$ is obtained from~\eqref{eq:js} by setting $F_s^>=0$ and 
$F_s^<=S_s^<$.
For simplicity we take phenomenological Gaussians with width $\Gamma_{\text{in}}$ centered 
around $E-U_s=I_s^{\rm in}$ and $T=0$ as source functions. 

Ideally, the injection energies $I_s$ would be the real ones, set by the ion's 
ionization energy, in case of resonant ionization, and the dielectric's electron affinity.
Both are usually a couple of $\mathrm{eV}$ away from the band edges. The relaxation 
and recombination kinetics, on the other hand, making at the end the space charge
inside the solid quasi-stationary with the plasma sheath, requires a resolution on the order 
of the phonon energy, which is typically $0.1\,\mathrm{eV}$. Resolving on that scale the 
whole energy range up to the actual injection points is computationally very expensive. 
To keep the numerical costs at an acceptable level, we move the injection energies $I_s$ 
below an energy cutoff dictated by numerical constraints. The principal mechanism of the 
model, relaxation and subsequent recombination of plasma-injected surplus charges inside 
the plasma-facing solid, remains intact.

\subsection{Numerical strategy}\label{subsec:numerics}

We now sketch the numerical approach employed to solve the transport problem, focusing
on the overall strategy to determine the various parameters required to selfconsistently
embed the double layer between field-free, quasi-neutral bulk regions. Technical details 
concerning the plasma side and the integration routines are relegated to 
appendices~\ref{app:plasma} and~\ref{app:inte}.

The plasma source issues at $U_c=U_w$ ions and electrons 
belonging to half-Maxwellians characterized by $T_{e,i}$ and $n_{e,i}^{\rm LM}$. 
Input parameters are only the temperatures. The densities are determined from the model
in a two-step procedure. First, enforcing the absence of an electric field and the 
quasi-neutrality at $U_c=U_p<U_w$, giving rise to the two conditions,
\begin{align}
        \mathcal{E}(U_p) &= 0~,\label{eq:condE(Up)}\\
        n(U_p) &= 0~,\label{eq:condn(Up)}
\end{align}
and combining them with the flux equality,
\begin{equation} 
        j_e(U_c) = j_i(U_c) ~,\label{eq:condj}
\end{equation}
the density ratio $\alpha=n_i^{\rm LM}/n_e^{\rm LM}$, to be interpreted as the strength 
of the plasma source, and the two potential parameters $U_p$ and $U_w$ 
can be determined. In a second step, the matching~\eqref{eq:matching} of the electric field 
across the interface, feeding in information form the solid side, is used to determine the 
absolute values of $n_e^{\rm LM}$ and $n_i^{\rm LM}$.
\begin{table}
\caption{The two sets of material parameters (a) and (b) we used in our numerical calculations.
For the Debye lengths the acceptor density $n_A$ is used instead of the intrinsic density
$n_{\rm int}$.  For set (b) only values different from set (a) are displayed.}
        \label{tab:solid}
        \begin{ruledtabular}
        \begin{tabular}{lll}
                                                  &       (a)      & (b)      \\
                \hline
                $E_g[\text{eV}]$                  &       1        &  2        \\
                $\hbar\omega_0[\text{meV}]$       &       75       &           \\
                $E_t[\text{eV}]$                  &       0.4      &  0.3      \\
                $\varepsilon$                     &       11.8     &           \\
                $\varepsilon_\infty$              &       12       &           \\
                $n_A[\text{cm}^{-3}]$             &       $10^{13}$  &  $10^{14}$   \\
                $n_{\rm int} [10^{10}\text{cm}^{-3}]$ &       4.922    &    $10^{-8}$    \\
                $N_t[\text{cm}^{-3}]$             &       $10^{20}$  &         \\
                $\sigma_s[\text{cm}^2]$           &       $10^{-15}$ &         \\
                $k_B T_{*,h} [\text{eV}]$         &       0.025    &           \\
                $m_{*,h}[m_e]$                    &       1        &           \\
                $\lambda_D^w[\mu\text{m}]$        &       1.821    &   0.576   \\
        \end{tabular}

\end{ruledtabular}
\end{table}

On the solid side, we use half-Maxwellians at $U_c=U_1$. 
The temperatures characterizing them are again input parameters, 
while the densities are determined by the absence of an electric field and the 
quasi-neutrality, yielding the three conditions 
\begin{align}
        \mathcal{E}(U_1) &= 0~,\label{eq:condE(U1)}\\
        n(U_1) &= 0~,\\
        n_h^\mathrm{LM}n_*^\mathrm{LM} &= n_{int}^2
\end{align}
with the intrinsic density
\begin{align}
        n_{\rm int} = \frac{1}{4}\left(\frac{k_B T_*}{\pi}\right)^{3/2}
        \left(m_*m_h\right)^{3/4}\exp \left( -\frac{E_g}{2k_B T_*} \right)~,
\end{align}
where we set $T_*=T_h$. From the three equations the two densities $n_*^{\rm LM}$ and 
$n_h^{\rm LM}$ as well as the band bending $U_1$ can be determined. The parameters 
of the source functions $S^<_{*,h}$ are either input parameters ($I_{*,h}^{\rm in}$, 
$\Gamma_{\rm in}$) or fixed by flux continuity ($n_{*,h}^{\rm in}$). All free parameters 
are thus determined and the double layer is selfconsistently embedded between the two 
quasi-neutral, field-free regions.

Due to the collisionality, the modeling on the solid side requires only one 
potential energy parameter, the band bending $U_1$. The distribution 
functions $F^\gtrless_{*,h}(U_c,E,T)$ can be taken as half-Maxwellians at 
$U_c=U_1$ because the vanishing of the field makes them in~\eqref{eq:BEQ} to 
annihilate the collision integrals. With half-Maxwellians, satisfying 
detailed balance, this can be enforced. On the plasma side, however, being 
collisionless, the distribution functions cannot be half-Maxwellians
at $U_c=U_p$. They have to be put in at $U_c=U_w>U_p$ by the Schwager-Birdsall
construction~\cite{SB90} leading to two potential energy parameters, $U_p$ and 
$U_w$.

In order to get the density and potential profiles to be employed in the 
embedding conditions just listed, the Boltzmann equation~\eqref{eq:BEQ} has to 
be solved. On the plasma side this can be done analytically. Following the 
approach of Schwager and Birdsall~\cite{SB90}, it leads to the expressions listed
in appendix~\ref{app:plasma}. Had we also included collisions there, a numerical
solution along the lines we now present for the solid side would be in order.

The numerical approach for solving the Boltzmann equations for electrons 
and holes inside the dielectric is an iterative scheme, originally proposed 
by Grinberg and Luryi~\cite{GL92} for transport problems where distribution 
functions are known at the two end points of the integration domain. 
It has proven its feasibility for solid-solid interfaces~\cite{DG94,DP98,KH02}
and can be based on a rewriting of the Boltzmann equation~\eqref{eq:BEQ} for 
right- and left-moving distributions in the form (in the following $s=*,h$)
\begin{align}\label{eq:sol>it}
F_s^{>}(U_c) &= \xi_s(U_c,U_c-\Delta) F_s^>(U_c-\Delta) \nonumber\\
& + \int\limits_{U_c-\Delta}^{U_c} \frac{\mathrm{d} U}{\mathcal{E}(U)} \frac{\Phi_s^>(U)}{v_s(U)} \xi_s(U_c,U)
\end{align}
and
\begin{align}\label{eq:sol<it}
F_s^{<}(U_c) &= \xi_s(U_c +\Delta,  U_c) F_s^<(U_c + \Delta) \nonumber\\
&+ \int\limits_{U_c}^{U_c + \Delta}   
\frac{\mathrm{d} U}{\mathcal{E}(U)} \frac{\Phi_s^<(U)}{v_s(U)} \xi_s(U,U_c)
\end{align}
with the integrating factor 
\begin{equation}\label{eq:I}
\xi_s(U_c,U_c') = \exp\left( -\int\limits_{U_c'}^{U_c} \frac{\mathrm{d} U}{\mathcal{E}(\bar
        U_c)}\frac{\gamma_s^\gtrless(U)}{v_s(U)} \right)~,
\end{equation}
where $\Delta$ is an arbitrary energy shift, but at the end it will be the 
basic discretization step in $U_c$-direction. The two 
equations are an exact rewriting of the original Boltzmann equations 
utilizing (i) the fact that in the variable $U_c$ they are ordinary 
first order differential equations and (ii) that the integrating factor 
$\xi_s(U_c,U_c')$ satisfies group properties. For brevity, the $E$ and $T$ 
dependencies of the various functions are again suppressed. 

The iteration scheme we employed for solving Eqs.~\eqref{eq:sol>it} and~\eqref{eq:sol<it} is 
illustrated in Fig.~\ref{fig:domain} of appendix~\ref{app:inte}. To obtain the distribution 
function $F^>_s(U_c)$ in the interval $U_1<U_c<0$, Eq.~\eqref{eq:sol>it} is iterated from 
$U_c=U_1$ to $U_c=0$, while $F^<_s(U_c)$ is obtained from~\eqref{eq:sol<it} by iterating it 
from $U_c=0$ to $U_c=U_1$, using at the starting points the boundary and matching conditions
specified above, and in the collision integrals the distribution functions of the previous
iteration loop. Special care has to be exercised by the discretization of the integrals
near singular points, where the left hand side of~\eqref{eq:BEQ} vanishes, leading to 
singularities in the integrals. Due to the collisions encoded in the functions 
$\gamma_s^\gtrless(U_c)$ and $\Phi_s^\gtrless(U_c)$ the variables $E$ and $T$ 
are not spectators of the integration procedure, as the simplified notation 
of~\eqref{eq:sol>it} and~\eqref{eq:sol<it} may suggest. In total, we have to iterate in a 
three-dimensional domain spanned by the variables $U_c$, $E$, and $T$. Further details 
and delicacies of the integration routines are discussed in appendix~\ref{app:inte}.

\section{Results}\label{sec:results}

\begin{table}[b]
\caption{Parameters of the source functions $S_*^<$ and $S_h^<$ describing the injection
of electrons and holes from the plasma into the solid. The width $\Gamma^\mathrm{in}$ 
is fine-tuned to match~\eqref{eq:matching} exactly even for the finite resolution of the
energy scales preventing the band bending $U_1$ to be specified to arbitrary
precision.}
\label{tab:source}
\begin{ruledtabular}
\begin{tabular}{lll}
$\Gamma^\mathrm{in}[\text{eV}]$ & $I_*^\mathrm{in}[\text{eV}]$ & $I_h^\mathrm{in}[\text{eV}]$  \\
        \hline
0.06                       &   0.2                   &  0.15                    \\
        \end{tabular}
\end{ruledtabular}
\end{table}

We now apply our model to a p-doped dielectric plasma-solid interface 
characterized by the parameters of Table~\ref{tab:solid}. Although we do 
not attempt to describe specific materials in contact with specific plasmas,
the parameters are chosen to represent a typical semiconductor facing 
a hydrogen plasma. The parameters of the source functions~\eqref{eq:Fin}, determined 
by the continuity of fluxes at $U_c=0$, are summarized in Table~\ref{tab:source} 
while the specifics of the plasma are given in Table~\ref{tab:plasma}. The values 
of the potential energy $U_c$ at $z=z_1, z_p, z_w$, denoting respectively 
the band bending, the sheath potential, and the drop of the sheath of the plasma 
source, listed in Table~\ref{tab:Uc}, are no input parameters. They arise from 
the selfconsistent matching of the solid and the plasma.   

\begin{table}[b]
\caption{Parameters of the collisionless hydrogen plasma in contact with the 
dielectrics (a) and (b) specified in Table~\ref{tab:solid}. For set (b) only 
values different from (a) are listed.}
\label{tab:plasma}
\begin{ruledtabular}
\begin{tabular}{lllllll}
system & $k_BT_e[\text{eV}]$ & $k_BT_i[\text{eV}]$  & $m_e[m_e]$ & $m_i[m_e]$ & $\lambda_D^p[\mu\text{m}]$  \\
        \hline
        (a) &     0.025 & 2              &   1      &  1836  & 9.107  \\
        (b) &           &                &          &        & 5.562
        \end{tabular}
\end{ruledtabular}
\end{table}

The trap density $N_t$ in Table~\ref{tab:solid} is artificially high because the 
coordinate transformation~\eqref{eq:U'}, mapping an infinite $z$-halfspace to a 
finite $U_c$-interval, restricts de facto the modeling to the region where the
band bending is significant. In general, this is favorable. But for the 
recombination process it is a problem since the recombination 
length $\lambda_R$, given in a rough approximation by 
\begin{align}
        \lambda_R \approx \frac{v_s}{\gamma_s^{trap}} \approx \frac{1}{\sigma_s N_t}~,
\end{align}
is for the realistic cross section $\sigma_s\approx 10^{-15}\,\text{cm}^{2}$ 
and the realistic trap density $N_t \approx 10^{16}\,\text{cm}^{-3}$ too large. 
It is on the order of $mm$ while the Debye length $\lambda_D^w$, setting
the scale of the space charge layer, and hence of our simulation domain, is 
only a few $\mu\text{m}$. To ensure complete recombination in the numerically 
resolved domain, necessary to prevent a pile-up of charges inside the solid, we 
have to increase thus $N_t$ by four orders of magnitude. 
\begin{figure}[t]
\includegraphics[width=\linewidth]{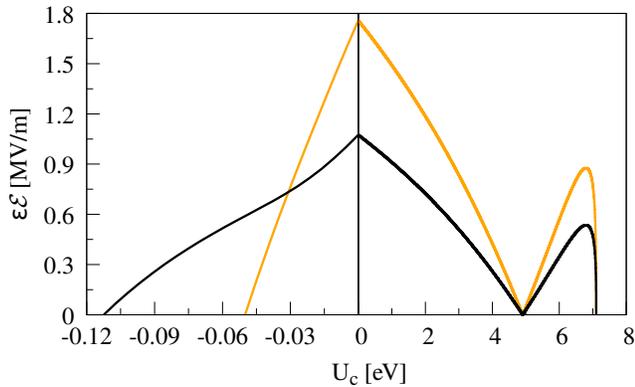}
\caption{Electric field multiplied by the dielectric function across the interface 
for the parameter sets (a) and (b), plotted respectively in black and orange,  
showing the matching~\eqref{eq:matching} to be satisfied.  
The maximum at $U_c \approx 6.8\,\text{eV}$ comes from 
the Schwager-Birdsall boundary condition. Since $U_p\approx 5\,\text{eV}$ 
effectively corresponds already to $z \approx \infty$ the maximum has no 
physical meaning. It is an artifact of implementing 
in a collisionless plasma a field-free, quasi-neutral bulk region.}
\label{fig:G}
\end{figure}

Let us start the discussion of the numerical data with the selfconsistent 
electric fields shown in Fig.~\ref{fig:G} for parameter sets (a) and (b). 
The matching condition~\eqref{eq:matching} is satisfied for both sets. 
Due to the higher acceptor concentration of set (b), the Debye length is 
shorter yielding a narrower space charge and a smaller band bending. The 
large fields on the solid side are in both cases mainly caused by the charge 
carriers due to the doping and not due to the surplus carriers coming from 
the plasma. On the plasma side, the field is due to the sheath in front of 
the solid. The Schwager-Birdsall boundary condition~\cite{SB90} leads 
to the non-monotonous behavior around $U_c\approx 6.8\,\text{eV}$. It is 
an artifact arising from the inflection point in the potential profile required 
to model in a collisionless plasma a field-free, quasi-neutral region representing 
the bulk plasma. Due to the coordinate transformation~\eqref{eq:U'}, the range 
of $U_c$ values shown in the plot corresponds essentially to two infinite 
halfspaces in the variable $z$. Mapping an infinite system to a finite one is  
an advantage of the change of coordinates. 
\begin{table}[b]
\caption{Numerical values of the selfconsistently determined potential energies at
$z=z_1$, $z=z_p$, and $z=z_w$ for the parameter sets (a) and (b) of 
Tables~\ref{tab:solid} and~\ref{tab:plasma}. Again, for (b) only values 
different from (a) are given.}
\label{tab:Uc}
\begin{ruledtabular}
\begin{tabular}{llll}
             system & $U_1[\text{eV}]$ & $U_p[\text{eV}]$ & $U_w[\text{eV}]$  \\
        \hline
                (a) & 0.1125          &   4.906    &  7.103              \\
                (b) & 0.05            &            &                     \\
        \end{tabular}
\end{ruledtabular}
\end{table}

Having found selfconsistent embeddings of the double layer, we now turn to the 
distribution functions for electrons and holes inside the solid. Without plasma
the electron and hole distribution functions are to a very good approximation 
half-Maxwellians, determined by the intrinsic carriers and the doping. Once the 
solid is in contact with the plasma, the distribution functions deviate from it 
due to the injection of carriers from the plasma and the band bending in response
to the sheath potential. Since the two parameters sets yield rather similar
results, we discuss below only data for one set.
\begin{figure*}[ht]
        \centering
         \subfloat[Electron distribution at $U_c=0$]{
                \includegraphics[width=0.46\textwidth]{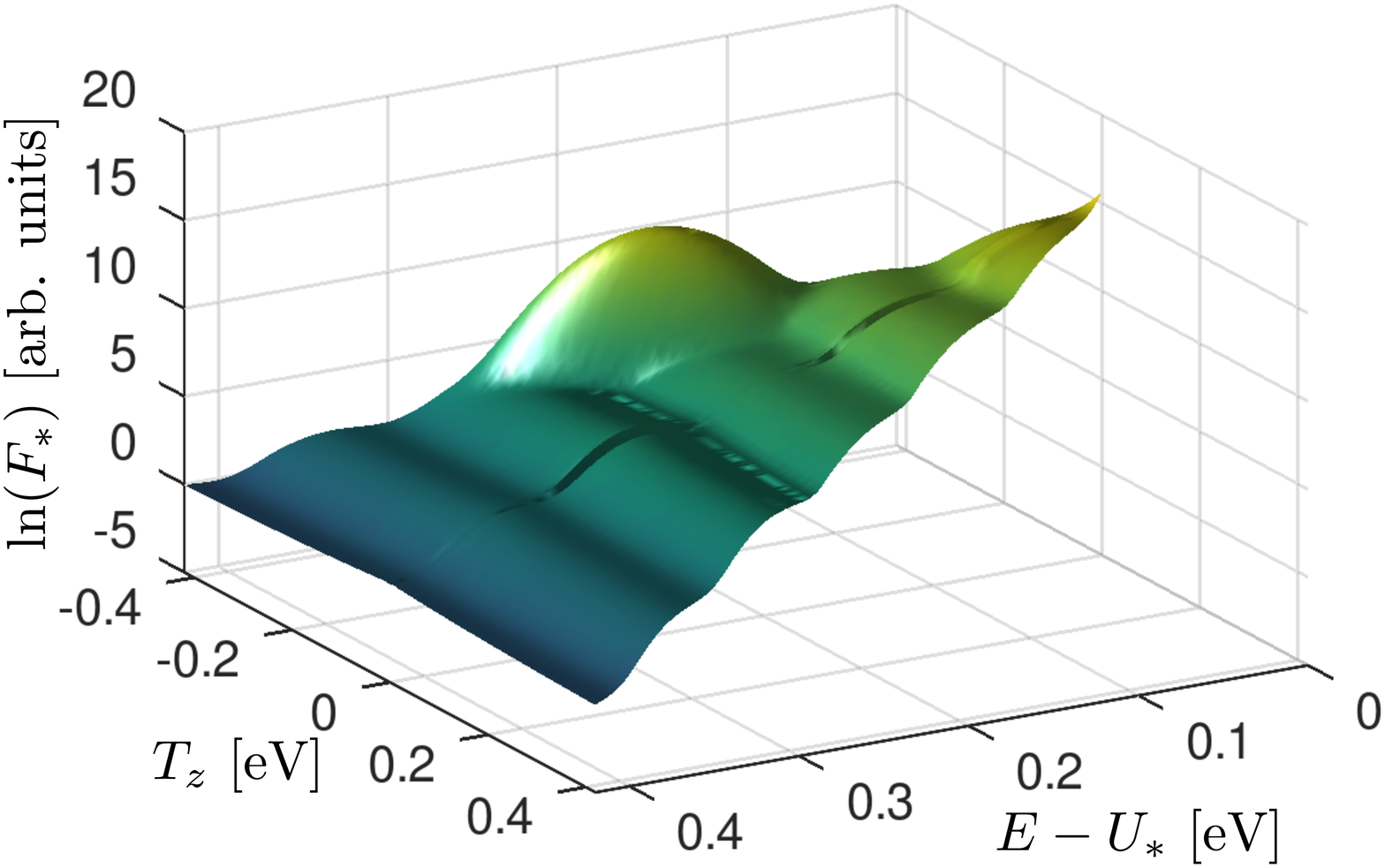}
                \label{fig:Fha}}
        \qquad
         \subfloat[Electron distribution at $U_c=21.4\,\mathrm{meV}$]{
                \includegraphics[width=0.46\textwidth]{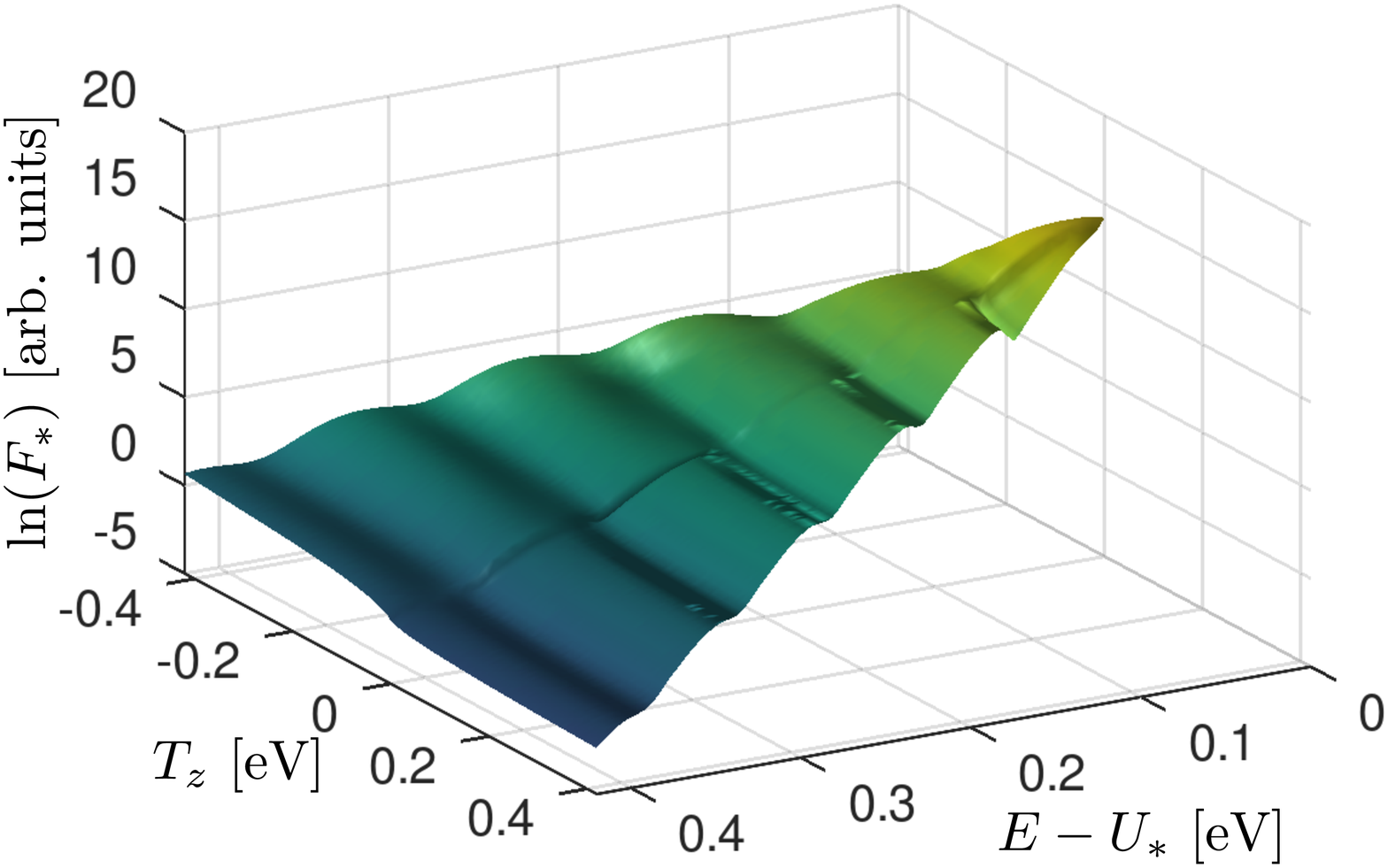}
                \label{fig:Fhb}}

         \subfloat[Hole distribution at $U_c=0$]{
                \includegraphics[width=0.46\textwidth]{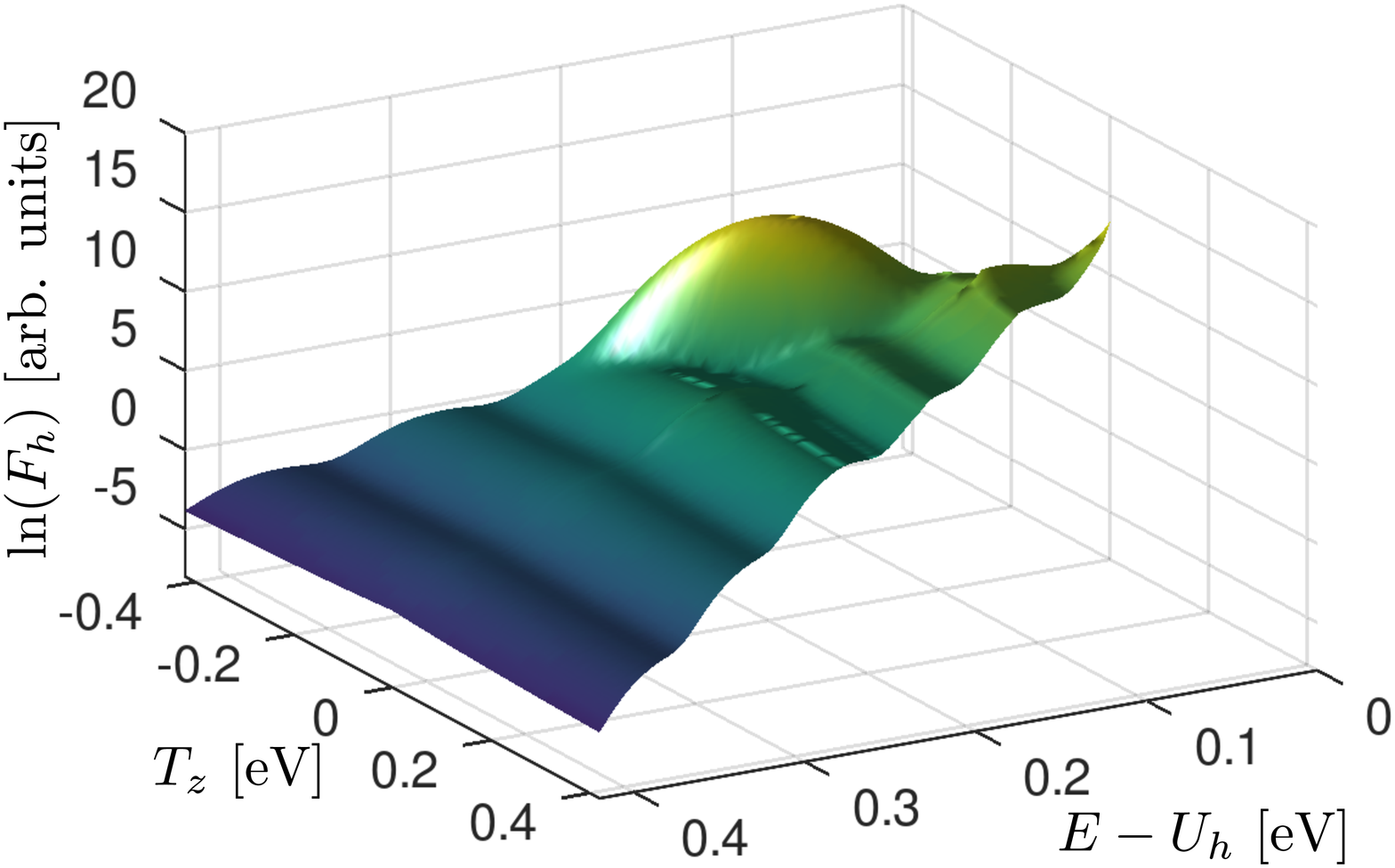}
                \label{fig:F*a}}
        \qquad
         \subfloat[Hole distribution at $U_c=21.4\,\mathrm{meV}$]{
                \includegraphics[width=0.46\textwidth]{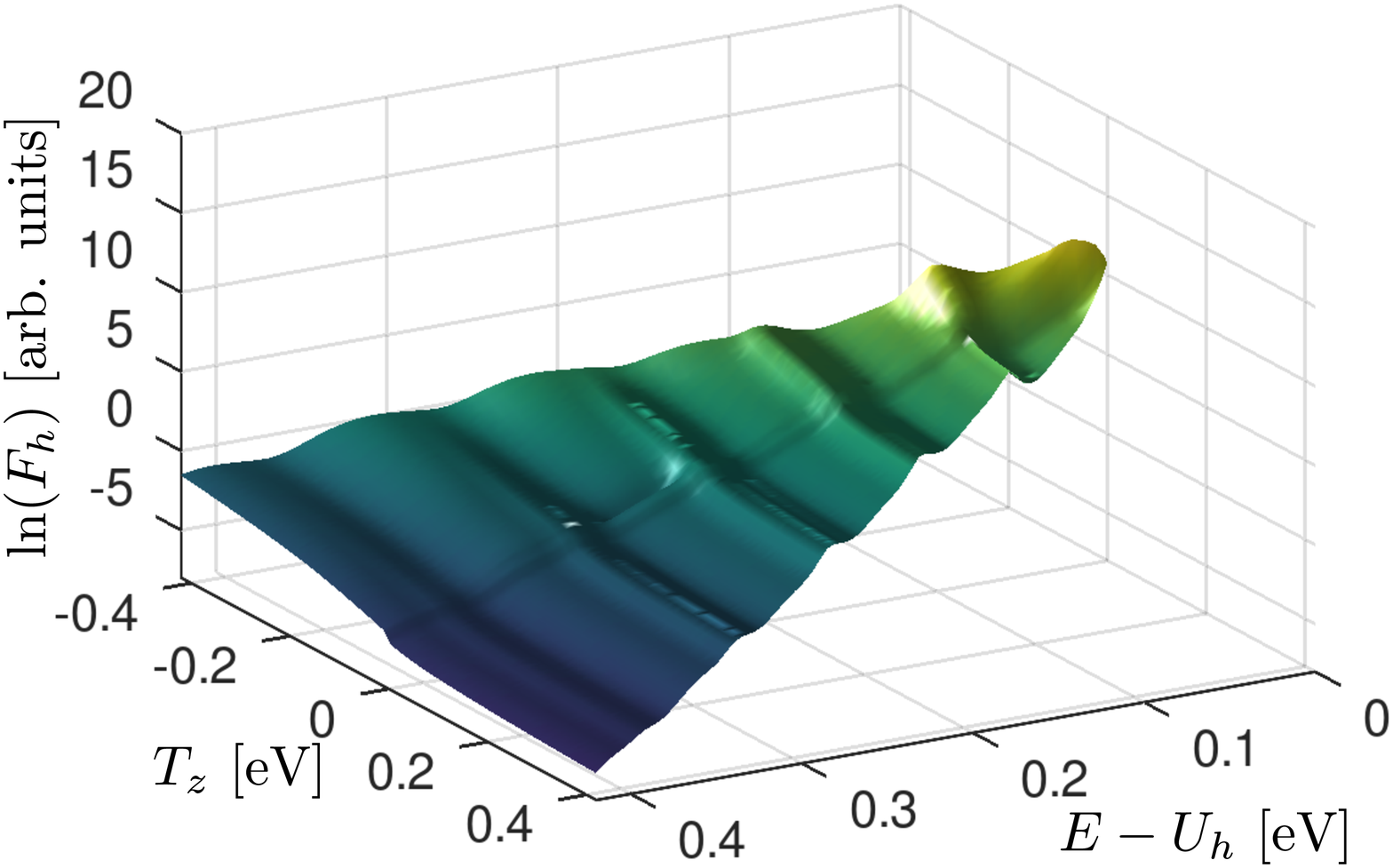}
                \label{fig:F*b}}
        \caption{Distribution functions for the injected electrons (upper panels) 
        and holes (lower panels) for parameter set (a) at $U_c=0$, that is, at the 
        interface and at $U_c=21.4\,\mathrm{meV}$. Instead of $E$ and $T$ the variables 
        $E-U_s$ and $T_z=E-U_s-T$ are used, where we attached to the latter a sign to 
        denote the distributions of left- ($T_z<0$) and right-moving ($T_z>0$) particles. 
        The injection peaks at $E-U_*=I_*^{\rm in}= 0.2\,\text{eV}$ and at 
        $E-U_h=I_h^{\rm in}= 0.15\,\text{eV}$ are clearly visible in the data for 
        $U_c=0$. Away from the interface, the peak gradually vanishes. Replicas
        due to phonon emission and absorption can be also seen, as well as the
        drop at $T_z=0$, signaling right-moving states to be less populated than 
        left-moving ones. Colors are used only for visibility reasons and the triangular
        shape is due to energy conservation.}
\label{fig:distributions}
\end{figure*}

Figure~\ref{fig:distributions} shows for parameter set (a) the deviations of the 
distribution functions from the half-Maxwellian background directly at the interface
at $U_c=0$ and inside the solid at $U_c=21.5\,\mathrm{meV}$. To visualize physical
effects more clearly, we plot $F^\gtrless_{*,h}$ for the two fixed values of $U_c$ 
as functions of $T_z=E-U_{*,h}-T$ and $E-U_{*,h}$, with $U_*=-\chi-U_c$, 
$U_h=U_c+E_g+\chi$, and $E$ running from $U_s$ to $E_{\rm max}=0.4\,\text{eV}$.
Besides the peak at $E-U_s=I_s^{\rm in}$ due to the source functions, clearly 
seen in the data for $U_c=0$, three further features can be identified: First, there is a 
series of peaks due to the scattering of the injected carriers on phonons. This 
is the energy and momentum relaxation of the carriers following injection from the 
plasma. Second, there is a step at $T_z=0$, separating the distributions for 
left- ($T_z<0$) and right-moving ($T_z>0$) carriers (encoded in the artificial sign of 
$T_z$). Because left-moving distributions have to be populated by backscattering events, 
which for interaction with optical phonons are rather unlikely, they are always smaller 
than the right-moving distributions populated by forward scattering. Third, the functions 
are maximal for $E-U_s\approx 0$ since the carriers accumulate at the band edges.

\begin{figure}
\includegraphics[width=\linewidth]{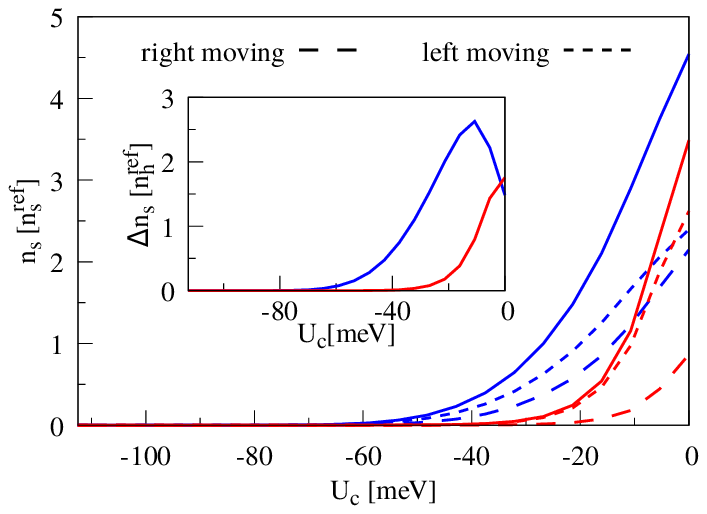}
\caption{Total (solid lines) and directional (long and short dashed lines) 
densities of the injected carriers for the parameter set (a). For plotting 
purposes we introduced reference densities $n_*^{ref}=6\cdot 10^{12}\,\text{cm}^3$ 
and $n_h^{ref}=10^{12}\,\text{cm}^3$. The positive difference of left and right moving 
densities $\Delta n_s$, shown in the inset, is on the same order of magnitude for 
electrons and holes indicating that for both polarities surplus carriers move 
more likely to the left than to the right. 
} 
\label{fig:density}
\end{figure}

That there are less right- than left-moving injected carriers can be also seen in
Fig.~\ref{fig:density}, where we plot for the parameter set (a) the directional 
electron and hole densities scaled to the reference densities given in the 
caption. The densities have been calculated from the distribution 
functions $F_{*,h}^\gtrless$ using~\eqref{eq:n} and subtracting from them the 
background densities due to the doping. All the surplus densities 
are maximal at $U_c=0$, that is, directly at the interface, and monotonously 
decrease to zero by approaching the bulk of the solid. In the inset the difference
of the densities of left- and right-moving carriers is shown. It is positive 
and of the same order for both polarities showing that both types of surplus carriers 
move preferentially to the left. From the plot we also see that injected 
electrons dominate injected holes as it should be for a double layer, where the 
positive, electron-depleted branch residing in front of the solid at $U_c>0$ 
has to be balanced by a net negative space charge inside the solid at $U_c<0$. 
The profiles demonstrate also that at quasi-stationarity the permanent influx 
of electrons and holes from the plasma does not lead to a pile-up of carriers inside 
the solid. Carrier recombination prevents this. 

\begin{figure}[t]
\includegraphics[width=\linewidth]{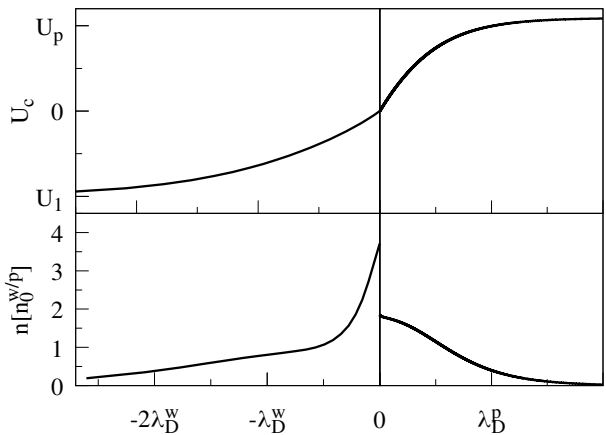}
\caption{Potential energy (upper panel) and net charge
density (lower panel) for parameter set (a) as a function of $z$ in units
of the Debye lengths, $\lambda_D^w=1.821\,\mu\text{m}$ and
$\lambda_D^p=9.107\,\mu\text{m}$. 
The kink in $U_c$ at $z=0$ signals the matching condition~\eqref{eq:matching}.
To fit the density profiles into a single plot, we scaled them on the solid 
side by $n_0^w=10^{13}\,\text{cm}^{-3}$ and on the plasma side by 
$n_0^p=-10^{12}\,\text{cm}^{-3}$.}
\label{fig:EDL}
\end{figure}

The net density and potential profiles of the double layer as a whole, embracing 
the solid and the plasma side, are shown in Fig.~\ref{fig:EDL} as a function of $z$ 
scaled to the corresponding Debye lengths. Since the results are similar for the 
two parameter sets we show again only data for set (a). Due to the difference
in the screening lengths, $\lambda_D^w\approx 1.8\,\mu\text{m}$ and 
$\lambda_D^p=9.107\,\mu\text{m}$ (see Tables~\ref{tab:solid} 
and~\ref{tab:plasma}), the charge neutrality of the double layer is not directly 
obvious but indeed satisfied due to the matching condition~\eqref{eq:matching}
which also gives rise to the different slopes of the potential energy profile for 
$z=0^-$ and $z=0^+$. The spatial scale of the double layer is set by the 
screening lengths. From the numerical values given in Tables~\ref{tab:solid}
and~\ref{tab:plasma} we see that for both parameter sets the width is on the
order of $1-10\mu\text{m}$ with the plasma side five- [set (a)] to ten-times 
[set (b)] thicker than the solid side. Notice the fast and slow decay of the 
density profile on the solid side in contrast to the more or less homogeneous decay 
on the plasma side. It indicates that electrons and holes do not recombine spatially 
concurrently in our model.

\begin{figure}[t]
\includegraphics[width=\linewidth]{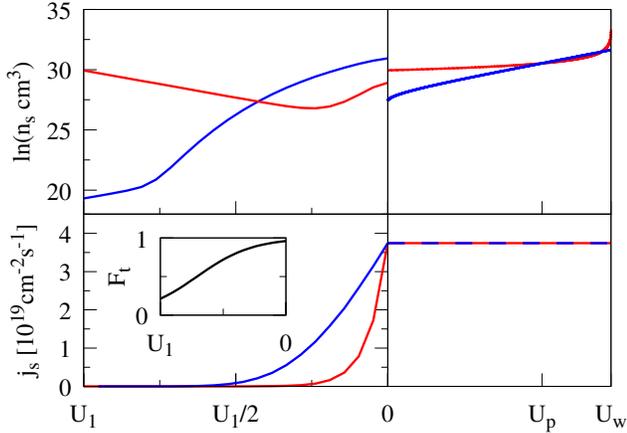}
\caption{
Mobile charge carriers (upper panel) and the fluxes
(lower panel) as a function of $U_c$ for parameter set (a). Data for electrons
(holes, ions) are plotted in blue (red). Inside the solid, electrons prevail 
for $U_1/2 < U_c < 0$, while for $U_c < U_1/2$ holes dominate, indicating the
p-doped bulk. The quasi-neutral bulk plasma emerges on the other side of the
interface for $U_c \approx U_p$. Also seen is the negative sheath in front
of the plasma source at $U_c=U_w$. The electron and ion fluxes of the plasma
merge at $U_c=0$ with the electron and hole fluxes of the solid, which then 
decay nonconcurrently due to the variation of the occupancy of the traps 
shown in the inset, placing empty (occupied) traps, required for electron 
(hole) recombination, further away from (closer to) the interface. 
}
\label{fig:merging}
\end{figure}

The merging of the double layer with the bulk regions on either side of the 
interface and the working of the recombination process are shown in 
Fig.~\ref{fig:merging}, where we plot, for both sides of the interface, as 
a function of $U_c$ the profiles of the carrier densities and fluxes. 
Recall, due to the coordinate transformation, the effectively infinite 
halfspaces in the spatial coordinate $z$ are mapped onto finite intervals
on the $U_c$ axis. The embedding can be clearly seen in the upper panel. 
On the solid side, only for $U_1/2 < U_c < 0$ is the electron density 
$n_*$ (blue) larger than the hole density $n_h$ (red), 
while for $U_c<U_1/2$ the ordering is reversed. Taking the acceptor density 
$n_A$ into account, which balances the hole density due to doping but is not 
included in the plot to make the scales comparable, the quasi-neutral p-doped 
region emerges for $U_c$ approaching $U_1$. On the plasma side, on the other 
hand, ions (blue) dominate electrons (red) for $0<U_c<U_p$, while for 
$U_c\approx U_p$ a quasi-neutral region appears merging, for $U_c$ approaching 
$U_w$, the negative sheath in front of the plasma source installed at $U_c=U_w$ 
by the Schwager-Birdsall construction. 

The workings of the recombination process can be inferred from the flux
profiles plotted in the lower panel of Fig.~\ref{fig:merging}. Electron 
and ion fluxes are equal on the plasma side and continuously merge at 
$U_c=0$ with the electron and hole fluxes. From the flux continuity 
$j_e(0)=j_i(0)=j_*(0)=j_h(0)$, the parameter $\alpha=n_i^{\rm LM}/n_e^{\rm LM}$,
introduced in appendix~\ref{app:plasma} and characterizing the strength of the 
plasma source can be obtained. For the data shown in Fig.~\ref{fig:merging} 
we find $\alpha=11$. Due to electron-hole recombination 
inside the solid the fluxes decay. The hole flux decays faster than the  
electron flux, indicating that holes are destroyed closer to the 
interface than electrons. That the recombination of holes and electrons
are spatially separated we have already noticed in the density profiles of
Fig.~\ref{fig:density}. It can be explained by looking at the trap occupancy shown 
in the inset of Fig.~\ref{fig:merging}. Holes have to recombine with an electron 
from the trap, that is, they require an occupied trap site, while electrons 
need empty trap sites. From the inset we see traps highly 
occupied close to the interface. Thus, in our model, holes preferentially 
recombine there, while electrons, requiring empty traps, have to move further 
into the solid, where the probability of finding them is higher. 

In reality, the two fluxes should decay equally fast. That in our model this is 
not the case is due to the approximation we used to determine the trap occupancy. 
Instead of the full electron and hole distribution functions, we employed in~\eqref{eq:Ft} 
only the half-Maxwellians arising from the doping background. By neglecting the 
contributions of the injected carriers, which are small but nevertheless present, 
the trap occupancy is not determined selfconsistently. In the present formulation 
of our model, it can thus not react to the injected carriers. Inserting the 
full distributions, however, would have led to nonlinear collision integrals, 
artificially dominating the kinetics due to the high trap densities $N_t$ we have 
to use to ensure complete recombination in the part of the simulation domain which 
is numerically resolved. Since the kinetic scenario we wanted to develop--destruction 
of plasma flux impinging on a dielectric by electron-hole recombination inside 
it--is not affected by the inconsistency, we did not include this additional complexity 
into the model.

\section{Conclusion}\label{sec:conclusion}
We have presented a selfconsistent kinetic model for the electric double layer at a 
dielectric plasma-solid interface that embraces plasma generation on one 
and plasma loss on the other side of the interface. Conduction band electrons and 
valence band holes are injected into the solid with unit probability for each 
impinging electron and ion. From the solid side, charge carriers cannot 
cross the interface which is thus modelled as a perfect absorber. Inside the 
solid electrons and holes scatter on optical phonons, leading to energy and 
momentum relaxation, before they recombine nonradiatively via traps in the energy 
gap of the dielectric. The microscopic picture encoded in our model is thus the one
of a plasma source whose fluxes are equalized and balanced by the recombination of 
electron and hole fluxes in the space charge region of the solid. 

Computational constraints in the numerical solution of the Boltzmann equation 
on the solid side forced us to treat charge injection by phenomenological source 
functions. The basic kinetics--injection of surplus charge carriers into the solid, 
followed by relaxation and recombination establishing a quasi-stationary double 
layer--is however still present in the simplified model. Based on an iterative scheme, 
geared towards solving Boltzmann equations with distribution functions specified 
at the endpoints of the integration domain, we presented the numerical solution 
of the kinetic equations, focusing in particular on the handling of singular 
points. A similar strategy could be applied on the plasma side in case it is made
collisional. 

Although quantitatively we cannot yet make hard predictions, because of the 
limitations of the phenomenological source functions, the perfect absorber assumption,
and the idealistic treatment of the electronic structure of the interface, which neglects,
for instance, adsorbate layers likely to be present in a plasma environment, the 
numerical results show the feasibility of the scheme. From the distribution functions of the 
holes and the electrons we calculated the solid-bound density and potential profiles merging 
the plasma sheath from the solid side. Combined with approaches describing the merging of 
the sheath with the bulk plasma in more detail than we have done, taking, for instance, 
ion-neutral collisions into account, a complete picture of the double layer can thus 
be developed. The results demonstrate moreover that the charge kinetics inside
the solid and the plasma can be treated on an equal footing, opening thus the 
door for a kinetic analysis of miniaturized semiconductor-based plasma devices 
combining gaseous and solid-state electronics.  

\section*{Acknowledgments}
Support by Deutsche Froschungsgemeinschaft through project BR-1994/3-1 is greatly acknowledged.

\appendix

\section{Collision integrals}\label{app:coll}
For the numerics it is convenient to split the collision integrals into an in-
and an out-scattering part,
\begin{equation}
\label{eq:Icoll}
I_\mathrm{coll}^\gtrless = \Phi^\gtrless_s(U_c,E,T) - \gamma^\gtrless_s(U_c,E,T) F^\gtrless_s(U_c,E,T)~,
\end{equation}
defining implicitly the functions $\Phi^\gtrless_s$ and $\gamma^\gtrless_s$ entering 
the Boltzmann equation~\eqref{eq:BEQ}. Starting with the standard forms of the 
collision integrals it is straightforward to work out expressions for $\Phi^\gtrless_s$ and 
$\gamma_s^\gtrless$. Below we give them for the scattering processes included in this work: 
electron ($s=*$) and hole ($s=h$) scattering by polar 
optical phonons~\cite{Ridley99} and nonradiative electron-hole recombination via traps in the 
energy gap of the dielectric according to the Shockley-Read-Hall mechanism~\cite{RJH16,Hall51,SR52}.

In the dilute limit, applicable to the situation we study, the collision integral for scattering 
by optical phonons becomes~\cite{Ridley99}
\begin{multline}
\label{eq:Iscat}
        I_s^\mathrm{scat}(z,\mathbf k) = \int\frac{\mathrm{d}^3k'}{(2\pi)^3} \bigg[ W(\mathbf k', \mathbf k)F_s(z,\mathbf k')\\
        - W(\mathbf k, \mathbf k')F_s(z,\mathbf k) \bigg]~,
\end{multline}
where the rate for scattering from $\mathbf{k}$ to $\mathbf{k'}$ is given by
\begin{multline}
\label{eq:W}
	W(\mathbf{k},\mathbf{k'}) = V\frac{2\pi}{\hbar}\abs{M(|\mathbf k-\mathbf k'|)}^2\\ 
	\times \big[ (1+n_b)\delta(E_{\mathbf {k}} - E_{\mathbf {k'}}-\hbar \omega_0) \\
	+ n_b\delta(E_{\mathbf {k}} - E_{\mathbf{ k'}}+\hbar \omega_0)\big]
\end{multline}
with   
\begin{equation}
 \label{eq:M}
 \abs{M(q)}^2 = \frac{\hbar \omega_0}{V}\frac{ 8\pi q^2\tilde{\alpha} \hbar c}{(q^2+q_s^2)^2} 
\left(\frac{\varepsilon_0}{\varepsilon_\infty} - \frac{\varepsilon_0}{\varepsilon}\right)
\end{equation}
the square of the matrix element for electron (hole)-phonon coupling. The standard notation 
is used throughout in the formulae, $E_{\mathbf {k}}$ is the kinetic energy of the electron 
(hole), $\hbar\omega_0$ is the energy of the (dispersionless, optical) phonon, and 
$n_b=1/(\exp(\hbar\omega_0/k_B T_*)-1)$ is the occupation number of the phonon. In the expression
for the matrix element, $c$ is the vacuum speed of light, $\tilde{\alpha}$ is the fine structure 
constant, $q_s$ is a screening momentum, and $\varepsilon$ and $\varepsilon_\infty$ are the dielectric 
constants at low and high frequencies, respectively. In the atomic units used in the main text, 
\begin{multline}
\label{eq:Wfull}
	W(\mathbf{ k},\mathbf {k'}) = 16 W_0\frac{q^2}{(q^2+q_s^2)^2}\\
	\times \big[(1+n_b) \delta(E_{\mathbf {k}} - E_{\mathbf {k'}}-\hbar \omega_0) \\+ n_b\delta(E_{\mathbf{ k}} - E_{\mathbf{ k'}}+\hbar \omega_0) \big]
\end{multline}
with $W_0=4\hbar \omega_0\left(1/\varepsilon_\infty-1/\varepsilon\right)$. Since the carrier 
concentrations are rather low, we neglect in the following the screening wave number $q_s$.
Parts of the calculations can then be performed analytically.

The functions $\gamma_s^\gtrless(U_c,E,T)$ and $\Phi_s^\gtrless(U_c,E,T)$ appearing 
in the Boltzmann equation~\eqref{eq:BEQ} are the integrals of either $W(\mathbf{k},\mathbf{k}')$ 
or $W(\mathbf{k}',\mathbf{k})F_s(z,\mathbf{k'})$ over $\mathbf{k}'$. In the limit $q_s=0$, 
we find, after rewriting the momenta $\mathbf{k}$ and $\mathbf{k}'$ in the coordinates 
$U_c, E, T, E'$, and $T'$, and distinguishing distributions for left- and right-moving particles,
\begin{widetext}
	\begin{equation}
		\label{eq:gamma}
	\gamma^\gtrless_s(U_c,E,T) = \frac{4W_0 }{v_s(U_c,E,0)}  
      \left[ n_b \arsinh\left( \sqrt{\frac{ E - U_s}{\hbar \omega_0 }} \right) 
      + (n_b + 1) \arsinh \left( \sqrt{\frac{E - U_s}{\hbar \omega_0} - 1}\right) \right]~,
	\end{equation}
	showing that this function is the same for both directions of motion, and 
	\begin{multline}
	\label{eq:Phi}
	\Phi_s^{\gtrless}(U_c,E,T) = W_0\times\\
	\sum\limits_\pm (n_b+\frac{1}{2}\pm \frac{1}{2} )\int\limits_{0}^{E-U_s\pm\hbar\omega_0}  
  \frac{\mathrm{d} T'}{v_s(U_s,E\pm\hbar\omega_0,T')} 
     \Bigg[\frac{ F_s^\gtrless(U_c,E\pm\hbar\omega_0,T') }{d_-(U_c,E,T,T', \pm\hbar\omega_0)}
	+\frac{F_s^\lessgtr(U_c,E \pm \hbar\omega_0,T')}{d_+(U_c,E,T,T',\pm\hbar\omega_0)}\Bigg]
	\end{multline}
	with
	\begin{equation}
	d_\pm(U_c,E,T,T', \hbar\omega_0) =\left[\left(T + T' + \left(\sqrt{E-U_s- T' 
       + \hbar\omega_0} \pm \sqrt{E-U_s-T}\right)^2 \right)^2 - 4TT'\right]^{1/2} ~~.
	\end{equation}
\end{widetext}
The second term of $\gamma_s^\gtrless$ and the in-scattering-by-absorption term in $\Phi_s^\gtrless$, 
that is, the term with the minus sign, only occurs for $E-U_s>\hbar\omega_0$. Note, the upper 
labels $\gtrless$ and $\lessgtr$ of the distributions on the right hand side of~\eqref{eq:Phi} are 
independent of the $\pm$ sign. They correspond to the $\gtrless$ of $\Phi_s^\gtrless$ 
on the left hand side of the equation.

We now turn to the kinetic formulation~\cite{RJH16} of the Shockley-Read-Hall 
electron-hole recombination~\cite{Hall51,SR52}. The collision integral coupling 
the trap occupancy $F_t$ with the electron distribution function $F_*^\gtrless$ reads
\begin{equation}
\label{eq:Itr*}
I_*^{{\rm tr}\gtrless}=\left(1 - F_*^\gtrless\right)\Gamma_G^* N_t F_t ~,
- F_*^\gtrless\Gamma_R^* N_t \left(1-F_t\right)
\end{equation}
while the one coupling $F_t$ it to the hole distribution function $F_h^\gtrless$ is
\begin{equation}
\label{eq:Itrh}
I_h^{{\rm tr}\gtrless}  = \left(1 - F_h^\gtrless\right)\Gamma_G^h N_t \left(1-F_t\right) 
- F_h^\gtrless\Gamma_R^h N_t  F_t~. 
\end{equation}
Therein $N_t$ is the trap density,
\begin{align}
\Gamma_R^s = \sigma_s v_s^\mathrm{tot}
\end{align}
is the recombination rate for species $s$, and
\begin{align}
\Gamma_G^* &= \sigma_* v_*^\mathrm{tot} \exp\left((E_t - (E-U_s+E_g))/k_BT_*\right)~,\\
\Gamma_G^h &= \sigma_h v_h^\mathrm{tot} \exp\left((-(E-U_s) - E_t)/k_BT_h\right)~
\end{align}
are the corresponding generation rates, where $\sigma_s$ is the capture cross section, 
$v_s^\mathrm{tot}$ is the total velocity (not to be confused with $v_s(E,U_c,T)$ which 
is the velocity in $z$-direction, that is, $v_s^\mathrm{tot} = v_s(E,U_c,0)$), and 
$E_t$ is the energy level of the traps. 

At quasi-stationarity, the trap occupancy is given by the detailed balance condition. 
Integrating~\eqref{eq:Itr*} and~\eqref{eq:Itrh} over $E$ and $T$ and equating the 
results, yields 
\begin{multline}
\label{eq:Ft}
F_t(U_c) = \bigg( m_* \int\frac{\mathrm{d} E \mathrm{d} T }{v_*} \Gamma_R^* F_*^{>+<} \\
+ m_h \int\frac{\mathrm{d} E \mathrm{d} T}{v_h}\Gamma_G^h\left(2-F_h^{>+<}\right) \bigg) \\
\times\bigg( m_* \int\frac{\mathrm{d} E \mathrm{d} T }{v_*} \left(\Gamma_R^* F_*^{>+<} +\Gamma_G^* 
\left(2 - F_*^{>+<} \right) \right)\\ 
+ m_h \int\frac{\mathrm{d} E \mathrm{d} T }{v_h} \left(\Gamma_R^h F_h^{>+<} + 
\Gamma_G^h \left(2 - F_h^{>+<} \right) \right) \bigg)^{-1}
\end{multline}
with $F_s^{>+<}$ denoting $F_s^>(E,U_c,T) + F_s^<(E, U_c,T)$. In the detailed balance 
condition~\eqref{eq:Ft}, we account only for the charge carriers due to the doping, described 
by half-Maxwellian distribution functions. The surplus electrons and holes coming from the plasma 
affect the balance only weakly because of their low density. In leading approximation, they 
can thus be neglected. 

Splitting~\eqref{eq:Itr*} and~\eqref{eq:Itrh} into out- and in-scattering contributions 
and distinguishing distributions for left- and right-moving electrons and holes yields  
\begin{align}
\gamma_*^{\text{tr}\gtrless} &=  \Gamma_G^* N_t F_t + \Gamma_R^* N_t \left(1-F_t\right)~,\\
\gamma_h^{\text{tr}\gtrless} &= \Gamma_G^h N_t \left(1-F_t\right) + \Gamma_R^h N_t  F_t~,\\
\Phi_*^{\text{tr}\gtrless} &=  \Gamma_G^* N_t F_t~,\\
\Phi_h^{\text{tr}\gtrless} &= \Gamma_G^h N_t \left(1-F_t\right)~,
\end{align}
which after eliminating $F_t$ by using~\eqref{eq:Ft} gives the form of the functions 
used in~\eqref{eq:BEQ}. 

\section{Plasma sheath}\label{app:plasma}
To make the present work self-contained, we summarize in this appendix the formulae for the
collisionless sheath forming on the plasma side of the interface. The merging of the plasma
sheath--in the absence of collisions--with the bulk plasma is established by a construction
due to Schwager and Birdsall~\cite{SB90}. It mimics the quasi-neutral, field-free bulk plasma
by an inflection point at $U_c=U_p$ arising between the sheath at the interface at $U_c=0$ and
the sheath in front of a plasma source imagined to sit at $U_c=U_w$. 

The plasma source at $U_c=U_w$ ejects electrons and ions with the half-Maxwellian
distributions~\eqref{eq:LM}. With the boundary condition $n_s(U_w)=n_s^\mathrm{LM}$ a
trajectory analysis of the collisionless Boltzmann equations on the plasma
side~\cite{BF17} leads to the density profiles
\begin{equation}
\label{eq:ni}
        \frac{n_i(U_c)}{n_i^\mathrm{LM}}  = f(a)-\sqrt{\frac{a}{\pi}}~,
\end{equation}
where $a=\frac{U_w - U_c}{k_B T_i}$, and
\begin{align}
\label{eq:ne}
        \frac{n_e(U_c)}{n_e^\mathrm{LM}}  = e^{-(a'+b)}\left[e^b - f(b) + \sqrt{\frac{b}{\pi}}\right]
\end{align}
with $a' = \frac{U_w-U_c}{k_B T_e}$ and $b = \frac{U_c}{k_B T_e}$. The function
\begin{equation}
\label{eq:f}
f(x) = \frac{1}{2}e^x \mathrm{erf}_c(\sqrt{x}) + \sqrt{x/\pi}
\end{equation}
is connected to the complementary error function $\mathrm{erf}_c(x) = 1-\mathrm{erf}(x)$. For 
the coordinate transformation~\eqref{eq:U'} we need the integrals over the profiles given by
\begin{equation}
\int\limits_{U_c}^{U_w}\mathrm{d} U \frac{n_i(U)}{n_i^\mathrm{LM}k_B T_i} = f\left(a\right)-\frac{1}{2}
\end{equation}
and
\begin{multline}
\int\limits_{U_c}^{U_w}\mathrm{d} U\frac{n_e(U)}{n_e^\mathrm{LM}k_B T_e} = 1-e^{-a'}\\
+ e^{a'+b}\left[ f(b) - f(a+b) \right]~.
\end{multline}

To determine the inflection point at $U_c=U_p$, the conditions $\mathcal{E}(U_p)=0$ and
$n(U_p)=0$ have to be worked out. Introducing $x_y = U_x/k_B T_y$ with $x\in \{p,w\}$
and $y\in\{i,e\}$ the condition of the vanishing electric field yields
\begin{equation}
\label{eq:G=0}
        \frac{n_i^\mathrm{LM}}{n_e^\mathrm{LM}} = \frac{k_B T_e}{k_B T_i}\frac{1-e^{p_e-w_e}
        + e^{-w_e}\left[ f(p_e) - f(w_e) \right]}{f\left(w_i-p_i\right)-\frac{1}{2} }~,
\end{equation}
whereas the vanishing of the net charge density becomes
\begin{multline}
\label{eq:n=n}
        \frac{n_i^\mathrm{LM}}{n_e^\mathrm{LM}} = \exp\left( p_e-w_e+p_i-w_e \right)\\
        \times\frac{1 + \Phi\left(\sqrt{p_e}\right)}{1-\Phi\left(\sqrt{w_i-p_i}\right)}~.
\end{multline}

Augmenting Eqs.~\eqref{eq:G=0} and~\eqref{eq:n=n} with the flux balance~\eqref{eq:condj},
using
\begin{equation}
\label{eq:je}
        j_e(U_c) = -n_e^\mathrm{LM}\sqrt{\frac{k_BT_e}{\pi}} e^{-w_e}
\end{equation}
and
\begin{equation}
\label{eq:ji}
        j_i(U_c) = -n_i^\mathrm{LM}\sqrt{\frac{k_B T_i}{m_i \pi}}~
\end{equation}
to be obtained upon inserting~\eqref{eq:ni} and~\eqref{eq:ne} into~\eqref{eq:js}, leads
finally to three equations for the four unknowns $U_w, U_p, n_e^{\rm LM}$, and $n_i^{\rm LM}$.
In the model of Schwager and Birdsall~\cite{SB90}, which does not include the solid, only three
of the parameters can thus be fixed. Considering the ratio $\alpha=n_i^{\rm LM}/n_e^{\rm LM}$
as the strength of the plasma source, $U_w, U_p$ and $\alpha$ are usually the parameters 
calculated. In our model, extending into the solid, the matching of the electric
field~\eqref{eq:matching} at $U_c=0$, that is, the charge neutrality of the double layer 
yields however an additional equation. At the end, we can thus determine all four parameters.

\section{Integration routines}\label{app:inte}
In this appendix we describe the integration routines used in the numerical treatment 
of Eqs.~\eqref{eq:sol>it} and~\eqref{eq:sol<it}, focusing on the discretization and 
the handling of singular points.
\begin{figure}
\includegraphics[width=\linewidth]{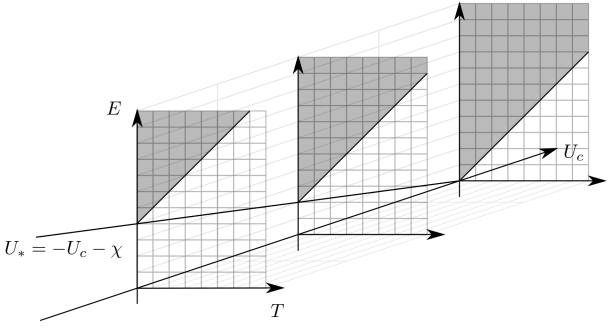}
\caption{Schematic illustration of the three-dimensional integration domain for 
conduction band electrons spanned by the variables $U_c$, $E$, and $T$. The 
domain is cut by the surface $v_*(U_c,E,T)=0$ leading to turning points at 
$T=E-U_*(U_c)$ separating the region where electrons are allowed to move shown 
in dark [$T<E-U_*(U_c)$] from the forbidden region [$T>E-U_*(U_c)$]. For the 
valence band holes the domain is divided by the function $T=E-U_h(U_c)$
giving rise to a different shape. The elementary discretization 
step, to be the same in all three dimensions, $\Delta=\hbar\omega_0/n$ with 
$\hbar\omega_0$ the phonon energy. The energy cutoff 
$E_{\bf k}^{\rm cutoff}=N \,\hbar\omega_0$ is used for the total kinetic 
energy $T+T_z$ of the carriers measured from the bottom of the bands, limiting 
thereby the variable $E$. For the results presented in this work we 
typically used $n=8$, $N=14$ yielding around $100^3$ discretization points.} 
\label{fig:domain}
\end{figure}

The three-dimensional integration domain, spanned by the variables $U_c$, $E$, and
$T$, is shown in Fig.~\ref{fig:domain}. For all three the discretization
step $\Delta$ is used, to be taken as a fraction of $\hbar\omega_0$. The potential 
energy $U_c$ ranges from $U_1$ to $0$, the total energy $E$ is at least $U_s$ and in 
principal unbound, and the lateral kinetic energy $T$ takes values from $0$ to $E-U_s$.
To keep the integration domain also in the variable $E$ finite, we use an energy 
cutoff $E_{\bf k}^{\rm cutoff}=0.4\,\mathrm{eV}$ for the total kinetic energy 
$T+T_z$ of the charge carriers measured from the bottom of the bands. 

By discretizing the integrals singular points have to be carefully treated. The 
square-root singularity due to the vanishing of $v_s(U_c,E,T)$ at $T=E-U_s$ can be 
removed by substitution. The $U_c$-integrals in Eqs.~\eqref{eq:sol>it} and~\eqref{eq:sol<it}
are then solved in one step, without further interpolation points, by linearization. 
Some integrals are however still singular because $\xi_s$ is an exponential function 
of a possibly diverging integral. They have to be done by hand. Two types of integrals
have to be distinguished: Integrals where $\mathcal{E}(U)\neq 0$ and integrals where 
$\mathcal{E}(U)=0$ which occur however only at the end point $U=U_1$.

First, we consider the case $\mathcal{E}(U)\neq 0$. The integrand in $\xi_s$ 
as well as the product of the functions in front of $\xi_s$, except of $v_s$, 
which has been removed by the substitution 
\begin{equation}
\label{eq:Z}
	Z = \sqrt{U_0\pm U}~,
\end{equation}
where the $+$ sign and $U_0=E-T+\chi$ is used for electrons, and the $-$ sign and 
$U_0=E-T-E_g-\chi$ is used for holes, can be linearized. Then, integrals of the form 
\begin{equation}
\int\limits_{Z_0}^{Z_1}\mathrm{d} Z \left(f_0 + Zf_1\right)\exp\left(-\left(g_0 + Z g_1\right)
\left(Z_1-Z\right)\right)
\end{equation}
appear for Eq.~\eqref{eq:sol>it}, while integrals of the type 
\begin{equation}
\int\limits_{Z_0}^{Z_1}\mathrm{d} Z \left(f_0 + Zf_1\right)\exp\left(-\left(g_0 + Z g_1\right)
\left(Z-Z_0\right)\right)
\end{equation}
are found for Eq.~\eqref{eq:sol<it}. The abbreviations, subsuming numerical coefficients 
arising from the linearization, should be clear from the context. For instance, 
$f_0=\Phi_s^\gtrless(U_c)/(\mathcal{E}(U_c)v_s(U_c))$. Using the identities 
\begin{equation}
\int \mathrm{d} x\exp\left(ax + bx^2\right) =
\frac{1}{2}\sqrt{\frac{\pi}{b}}\exp\left(-\frac{a^2}{4b}\right)\mathrm{erfi}\left(\frac{a+bx}{2\sqrt{b}}\right)
\end{equation}
and
\begin{multline}
\int \mathrm{d} x x \exp\left(ax + bx^2\right) =
\frac{1}{2b}\exp\left( ax + bx^2 \right)\\
-\frac{1}{4}\frac{a}{b}\sqrt{\frac{\pi}{b}}\exp\left(-\frac{a^2}{4b}\right)\mathrm{erfi}
\left(\frac{a+bx}{2\sqrt{b}}\right)~,
\end{multline}
the integrals can be related to the imaginary error function $\mathrm{erfi}(x)$
to be calculated as follows: For positive $b$ we use the 
Dawson function $F(x)=\sqrt\pi\exp(-x^2)\mathrm{erfi}(x)/2$ while for negative 
$b$ we employ the identity $\mathrm{erfi}(\mathrm{i} x)=\mathrm{i}\,\mathrm{erf}(x)$.
The error function $\mathrm{erf}(x)$ in turn is evaluated by routines of standard 
libraries. For large arguments, where the routines have problems, we expand
$\mathrm{erf}(x)$ together with the factor $\exp\left(-a^2/4b\right)$ 
into a power series. 

To deal with the integrals where $\mathcal{E}=0$, we assumed and verified 
a posteriori, that $\mathcal{E}$ starts linearly with $U_c$. The divergence of
$1/\mathcal{E}$ in the integral of the exponent of $\xi_s$ is then cancelled by the
same divergence in the $U_c$-integrals of~\eqref{eq:sol>it} and~\eqref{eq:sol<it}. 
A linear approximation for $\mathcal{E}(U_c\gtrsim U_1)$ together with~\eqref{eq:Z}
implies $\mathcal{E}(Z)\sim Z^2-Z_0^2$. Linearizing 
the remaining parts of the integrands relates the integrals to the incomplete 
beta function, 
\begin{equation}
\label{eq:beta}
	B_x(\alpha+1,1-\beta)=\int\limits_{0}^{x}\mathrm{d} y\frac{y^\alpha}{\left(1-y\right)^\beta}~,
\end{equation}
or--in case $Z_0=0$ (that is, for $T=E-U_s$)--to the incomplete gamma function
\begin{equation}
\Gamma(a,x) = \int\limits_0^{1/x}\mathrm{d} y y^{-a-1}\exp\left(-\frac{1}{y}\right)~.
\end{equation}
The parameters $a,\alpha,\beta$ and $Z_0$, again numerical coefficients arising from the 
linearization, can be straightforwardly albeit tediously determined. Depending on the 
arguments, the beta function is evaluated either in terms of a continued fraction 
representation or in terms of the hypergeometric function, using
\begin{align}
B_x(a,b) = \frac{x^a}{a}\prescript{}{2}F_1(a,1-b,a+1,x)~.
\end{align}
Likewise, the incomplete gamma function is obtained from a continued fraction expansion in 
cases where the evaluation with routines from standard libraries fails. 

After the integrals have been evaluated in the form just described we have an algebraic set 
of equations which can be iterated in the three-dimensional domain shown in 
Fig.~\ref{fig:domain}. The particular shape of the domain depends on the species through 
the function $U_s$. We found convergence to be reached faster if the iteration process does 
not destroy detailed balance in the phonon collision integrals. We thus put--for 
phonon collisions only--in~\eqref{eq:sol>it} and~\eqref{eq:sol<it} the term
$-\gamma_s^\gtrless F_s^\gtrless$ into the function $\Phi^\gtrless_s$.

\bibliography{ref}

\end{document}